\documentclass[a4paper,12pt]{article}

\usepackage{float}
\usepackage{harvard}

\citationmode{abbr} \citationstyle{dcu}
\usepackage{amsmath}
\usepackage{amssymb}
\usepackage{graphicx}
\usepackage{graphicx}
\usepackage{lscape}
\usepackage{epic}

\begin{document}


\title{The role of optimization on the human dynamics of tasks execution}
\author{Daniel O. Cajueiro and Wilfredo L. Maldonado}
\date{}
\maketitle

\begin{center}Department of Economics, Catholic University of Brasilia, 70790-160,
Brasilia, DF, Brazil.\end{center}

\begin{abstract} In order to explain the empirical evidence that
the dynamics of human activity may not be well modeled by Poisson
processes, a model based on queuing processes were built in the
literature~\cite{bar05}. The main assumption behind that model is
that people execute their tasks based on a protocol that execute
firstly the high priority item. In this context, the purpose of
this letter is to analyze the validity of that hypothesis assuming
that people are rational agents that make their decisions in order
minimize the cost of keeping non-executed tasks on the list.
Therefore, we build and solve analytically a dynamic programming
model with two priority types of tasks and show that the validity
of this hypothesis depends strongly on the structure of the
instantaneous costs that a person has to face if a given task is
kept on the list for more than one step. Moreover, one interesting
finding is that in one of the situations the protocol used to
execute the tasks generates complex one dimensional dynamics.

\end{abstract}

\section{Introduction}

Empirical evidence has shown that the dynamics of inter-event
times driven by human actions may not be random and not well
approximated by Poisson processes~\cite{paxflo96,masmon03,sca06}.
Based on this, Barab\'{a}si~\cite{bar05} developed a very
interesting model of human activity where the distribution of the
inter-event time is a consequence of a decision queue processes.
He considers that among the most relevant protocols for driving
human dynamics, e.g. first-in-first-out protocol, random protocol
and a protocol based on the execution of the high priority item,
this later protocol seems to be the most important. In this
protocol, while high priority tasks are executed as soon as they
are added to the list, low priority tasks wait for a long time
until all high priority tasks are executed, i.e., the instants of
execution of low priority tasks are separated by long times of
inactivity. Using this assumption, it is numerically~\cite{bar05}
(analytically~\cite{vaz05})  shown that the distribution of
inter-event times follows a power law.

Two interesting contributions were introduced by~\cite{grilin06}.
First the authors map the variable length priority model
considered above onto a model of biased diffusion deriving
asymptotic distributions for the inter-event times. Second, in
order to investigate the arising of power laws in more general
situations, they generalize the fixed length model queue to
contain tasks with a priority label and with a class label where
there is always an active class and an inactive class. If the
highest priority task of the inactive class exceeds that of the
active class by at least a fixed switching cost, the inactive
class becomes active and the active class becomes inactive.

An interesting discussion is considered in~\cite{ken06,baroli06}
where it is argued that other mechanisms contribute for the
distributions of waiting times such as deadlines, time dependence
of priorities and the social context of the problem. In line with
this debate, \cite{bla06} relaxes the assumption that the
priorities of tasks do not change over time and studies queueing
systems where deadlines are assigned to the incoming tasks and the
urgency to attend a task increases with time showing that only in
the former
 model fat tails arise naturally as consequence of the scheduling
 rule.

In this letter, we investigate the assumption that people execute tasks on
a protocol that execute firstly the high priority item. In
particular, we suppose that people assign priorities to the tasks
on their lists in order to minimize some cost index, i.e., a cost
associated to the fact of not processing a given collection of
tasks in a given time step. Therefore, based on this assumption
and inspired on~\cite{bar05,vaz05,grilin06}, we have built a
discounted stochastic dynamic programming model with two types of tasks
(low and high priority tasks) and a cost per stage for keeping a
number of low and high priority tasks without processing.

This is not the first time that a kind of optimization principle
is used to understand the structure and dynamics of complex
systems. In~\cite{rodrin92,caj05}, for instance, it is shown that
complex networks may arise from optimization principles.

It is also important to stress that although there is a large
literature dealing with control of queue
discipline~\cite{cra77,kitryk95,sen99} which this work is related,
the model presented in this letter is neither an extension nor a
particular case of any of these results.

We have found that the type of protocol used to execute tasks is
strongly dependent on the kind of instantaneous cost of keeping a
task in the queue for an additional stage. When linear costs are
used the protocol of executing preferentially the high priority
costs always is the best solution. However, this does not happen
when quadratic costs are considered. In this case different types
of protocol are considered. Furthermore, depending on the
parameters of the system, the protocol considered generates
complex one dimensional dynamics.

\section{Setup of the problem} We consider that there are two queues waiting for a service on a single server.
Let $g(x_L,x_H)$ be the current cost of having state $(x_L,x_H)$
which is the state of the system, $x_{L}$ ($x_H$) is the number of
tasks in the first (second) queue. We say that the first queue is
a low priority queue (or the second queue is a high priority
queue) if $\frac{\partial g(x_L,x_H)}{\partial
x_L}|_{x_L=x_H}<\frac{\partial g(x_L,x_H)}{\partial
x_H}|_{x_L=x_H}$. We assume that this is the case. The dynamics of
these queues are modeled as follows: At each discrete time step
with probability $\lambda\rho$ a new task arrives in the queue
formed by high priority tasks and with probability $\lambda
(1-\rho)$ a new task arrives in the queue formed by low priority
tasks. Within each of the queues the tasks are executed on a
First-In, First-Out basis. With probability $\mu u(x_L,x_H)$ the
first task of the high priority queue is executed and with
probability $\mu (1-u(x_L,x_H))$ the first task of the low
priority queue is executed. We assume here that $u(x_L,x_H)$ is a
state dependent control variable that the agent will choose in
order to minimize the total cost function
$J_u(x_L,x_H)=E_{x_L,x_H}^{u}[\sum_{t=1}^{\infty}\alpha^t
g((x_L(t),x_H(t)))]$, where $E_{x_L,x_H}^{u}[\cdot]$ is the
expected value conditioned to the current state $(x_L,x_H)$ and
the state control variable $u$ and $\alpha$ is the discount
factor.

Due to the principle of optimality~\cite{bel57,ber01} and the
Banach fixed point theorem, if the minimum cost function
$J(x_L,x_H)=\min_{u(x_L,x_H)\in [0,1]}J_u(x_L,x_H)$ exists, it
must be given by the unique solution of the Bellman equation, that
may be written as
\begin{equation}J(x_L,x_H)=F(x_L,x_H)+\min_{u(x_L,x_H)\in [0,1]}u(x_L,x_H)G(x_L,x_H)\label{bellmanEquation} \end{equation}
where
\begin{eqnarray}F(x_L,x_H)&=&g(x_L,x_H)+ \lambda  \rho  (1-\mu)  [\alpha
  J(x_L,x_H+1)]\nonumber\\
  &+&\lambda  (1-\rho)  (1-\mu)  [\alpha J(x_L+1,x_H)]\nonumber\\
  &+&(1-\lambda)  \mu  [\alpha  J(x_L-1,x_H)]\nonumber\\&+&\rho  \lambda  \mu  [\alpha  J(x_L-1,x_H+1)]\nonumber\\
  &+& (1-\rho)  \lambda  \mu  [\alpha  J(x_L,x_H)]\nonumber\\
  &+& (1-\lambda)  (1-\mu)  [\alpha  J(x_L,x_H)] \end{eqnarray}

and

\begin{eqnarray}&&G(x_L,x_H)=\nonumber\\&&(1-\lambda)  \mu
[\alpha
    (J(x_L,x_H-1)-J(x_L-1,x_H))]\nonumber\\
    &+&  \rho  \lambda  \mu  [\alpha
    (J(x_L,x_H)-J(x_L-1,x_H+1))]\nonumber\\
    &+& (1-\rho)  \lambda  \mu  [\alpha
    (J(x_L+1,x_H-1)-J(x_L,x_H))] \label{eq:G}\end{eqnarray}

Since the optimization problem (\ref{bellmanEquation}) is a linear
programming problem, the optimal control $u(x_L,x_H)$ in each state
$(x_L,x_H)$ will depend explicitly on the signal of $G(x_L,x_H)$.
If $G(x_L,x_H)>0$, then $u(x_L,x_H)=0$. If $G(x_L,x_H)<0$, then
$u(x_L,x_H)=1$. Finally, if $G(x_L,x_H)=0$, $u(x_L,x_H)$ is a
mixed strategy that may present any value in the interval $[0,1]$.
It is quite intuitive this result. Indeed, one may note that the terms in square brackets defined in
$G(x_L,x_H)$, equation (\ref{eq:G}), comprises the variations in the cost function due to changes in the states of the queue related to the execution of one of the tasks.

Since the properties of the solution of $J(x_L,x_H)$ of the
Bellman equation (\ref{bellmanEquation}) are strongly dependent on
choice of the cost per stage $g(x_L,x_H)$, in the next sections,
two different choices for $g(x_L,x_H)$ are investigated.

\section{Linear costs}

In this section, we assume that $g(x_L,x_H)=h_L x_L+h_H x_H$, for
$0<h_L<h_H$, i.e., the current cost of having one additional high
priority task in the queue is larger than having one additional
low priority task in the queue.

Since the space of polynomials of degree 1 with sup-norm is a
Banach space, one can show inductively, making recursive
iterations of the dynamic programming mapping, that $J(x_L,x_H)$
is also linear. Therefore, for $x_L>0$ and $x_H>0$, guessing this
form, one may easily solve the Bellman equation
(\ref{bellmanEquation}) and show that the cost function is given
by \footnote{The constants are given by
$c_L=\frac{h_L}{1-\alpha}$, $c_H=\frac{h_H}{1-\alpha}$ and
$c=\frac{\alpha}{(1-\alpha)^2} (\lambda (1-\rho)
h_L+(\lambda\rho-\mu)h_H )$.}
\begin{equation}J(x_L,x_H)=c+c_L x_L +c_H x_H\end{equation}

\vspace{2mm}

Furthermore,
\begin{equation}G(x_L,x_H)=\mu \frac{\alpha}{1-\alpha} (h_L - h_H) \end{equation}
is always negative implying that $u(x_L,x_H)=u=1$ for every state
$(x_L,x_H)$. Therefore, if linear costs are considered, the
protocol to be considered is the one based on the execution of the
high priority task whenever there is at least one item in this
queue, i.e., $x_H>0$. This kind of protocol was very well studied
in~\cite{bar05,vaz05,grilin06} where analytic results for the
emerging of power laws may be found. In the next section, a much
wealthier situation happens where the optimal policy is not only
limited to execute the high priority item in the queue, but the
optimal policy is state-dependent.

\section{Quadratic costs} Now, we assume that $g(x_L,x_H)=h_L x_L^2+h_H
x_H^2$, for
$0<h_L<h_H$. Following the same reasoning already presented before for the
linear cost case, one may conclude a quadratic form for the cost
function.

Solving the Bellman equation, one may show that the solution of
the problem depends explicitly on the signal of the function
$G(x_L,x_H)$, defined in (\ref{eq:G}), in the state $(x_L,x_H)$.
In fact, three different regions will arise. We will call region
$A$ the domain of $(x_L,x_H)$ where $G(x_L,x_H)>0$, region $B$ the
domain of $(x_L,x_H)$ where $G(x_L,x_H)=0$ and region $C$ the
domain of $(x_L,x_H)$ where $G(x_L,x_H)<0$.

We have found that the minimum cost function is given by

\begin{equation}J(x_L,x_H)=\left\{\begin{array}{c}
                                    J^A(x_L,x_H)\;\mathrm{if}\; (x_L,x_H)\in A \\
                                    J^B(x_L,x_H)\;\mathrm{if}\; (x_L,x_H)\in B \\
                                    J^C(x_L,x_H)\;\mathrm{if}\; (x_L,x_H)\in C \\
                                  \end{array}\right.\label{J} \end{equation}
and the optimal control is given by

\begin{equation}u(x_L,x_H)=\left\{\begin{array}{c}
                                    0\;\mathrm{if}\; (x_L,x_H)\in A \\
                                    u\in [0,1]\;\mathrm{if}\; (x_L,x_H)\in B \\
                                    1\;\mathrm{if}\; (x_L,x_H)\in C \\
                                  \end{array}\right.\label{u} \end{equation}

where for $i=A,B,C$
\begin{eqnarray}J^i(x_L,x_H)&=&c^{i}+c_{L1}^{i}x_L + c_{H1}^i x_H +
c_{L2} x_{L}^{2} + c_{H2}x_{H}^{2}\nonumber\\\end{eqnarray}

\begin{eqnarray}G^i(x_L,x_H)&=&\mu[c_{L1}^{i} -c_{H1}^{i}+2 (c_{L2} x_l -c_{H2} x_H)\nonumber\\
&+& c_{L2} (2\lambda (1-\rho) -1)
+c_{H2}(1-2\lambda\rho)]\nonumber\\
\end{eqnarray}
 and $G(x_L,x_H)=G^i(x_L,x_H)$, if $(x_L,x_H)\in i$~\footnote{The constants are given by $c_{L2}=\frac{h_L}{1-\alpha}$,
$c_{H2}=\frac{h_H}{1-\alpha}$, $c_{L1}^{A}=\frac{2\alpha h_L (-\mu
+ \lambda (1-\rho))}{(1-\alpha)^2}$,
$c_{L1}^{B}(\delta)=\frac{1}{(h_L+h_H)}\{\frac{h_{L}^{2}}{(1-\alpha)}[1-
2\lambda(1-\rho)]
 +\frac{h_L h_H}{(1-\alpha)^2} [2\alpha(\lambda-\mu)
+(2\lambda\rho-1+2\delta)(1-\alpha)]$, $c_{L1}^{C}=\frac{2\alpha
h_L \lambda (1-\rho)}{(1-\alpha)^2} $, $c_{H1}^{A}=\frac{2\alpha
h_H \lambda \rho}{(1-\alpha)^2}$,
$c_{H1}^{B}(\delta)=\frac{1}{(h_L+h_H)}\{\frac{h_{H}^{2}}{(1-\alpha)}[1-
2\lambda\rho-2\delta] +\frac{h_L h_H}{(1-\alpha)^2}
[2(\lambda-\mu\alpha) -(2\lambda\rho+1)(1-\alpha)]\}$,
$c_{H1}^{C}=\frac{2\alpha h_H (-\mu+\lambda\rho)}{(1-\alpha)^2} $,
$c^A=\frac{\alpha}{(1-\alpha)^3}\{[(1-\alpha)(\mu+(1-\rho)\lambda))+2\alpha(\lambda^2+\mu^2
-\mu\lambda)+2\rho\lambda(\mu(1+\alpha)+\lambda\alpha(\rho-2))-2\mu\lambda]h_L
+ \lambda\rho[1-\alpha+2\lambda\rho\alpha]h_H$,
$c^B=\frac{\alpha}{(h_L+h_H)(1-\alpha)^3}\{[2\lambda\alpha(1-\alpha)(-\rho^2
\lambda+2\lambda\rho-1-\lambda-\rho) ]h_{L}^2
+[\delta(2\lambda\rho\alpha^2+4\lambda\alpha-4\lambda\rho\alpha+1-2\lambda-2\alpha+\alpha^2-2\lambda\alpha^2+2\lambda\rho)+2\lambda(-\mu\alpha+2\lambda\rho^2\alpha^2+2\lambda\rho\alpha
-2\lambda\rho^2\alpha-\alpha^2
\mu-2\lambda\alpha^2\rho+\lambda\alpha^2)+2\mu(\alpha-\alpha^2
+\mu\alpha^2)]h_L h_H
+[\delta(-4\lambda\rho\alpha+2\delta-\alpha^2+2\delta\alpha^2+2\lambda\rho\alpha^2
-1
-4\alpha\delta+2\alpha+2\lambda\rho)+2\lambda\rho(\lambda\rho\alpha^2-\lambda\rho\alpha-\alpha^2+\alpha)]h_{H}^2
\}$ and $c^C=\frac{\alpha}{(1-\alpha)^3}\{[\lambda
(1-\alpha)(1-\rho)+2\lambda^2\alpha (1-\rho)^2]h_L
+[(1-\alpha)(\mu+\lambda\rho)+2\mu^2\alpha +2\lambda\rho
(\lambda\rho\alpha -\mu\alpha -\mu]h_H \}$.}.

\vspace{2mm}

Furthermore, the parameter
$\delta\in[\underline\delta,\overline\delta]$, defines the set of
points of $\Re^2$ such that
$G(x_L,(h_L/h_H)x_L+\delta)=0$~\footnote{The constants
\underline{$\delta$} $\,$ and $\overline\delta$ are respectively
given by
\underline{$\delta$}$=\frac{1}{1-\alpha}\{[\frac{1-\alpha}{2}-\lambda\rho]+
\frac{h_L}{h_H}[-\frac{1-\alpha}{2}+\lambda(1-\rho)-\alpha\mu]\}$
and
$\overline{\delta}=\frac{1}{1-\alpha}\{[\frac{1-\alpha}{2}-\lambda\rho+\alpha\mu]
+ \frac{h_L}{h_H}[-\frac{1-\alpha}{2}+\lambda(1-\rho)]\}$.}.

Indeed,
\begin{equation}\delta\rightarrow \underline{\delta} \Rightarrow \left(J^B(\delta)\rightarrow J^A,\;G^B(\delta) \rightarrow G^A \right)\end{equation}
and
\begin{equation}\delta\rightarrow \overline{\delta} \Rightarrow \left(J^B(\delta)\rightarrow J^C,\;G^B(\delta) \rightarrow G^C \right)\end{equation}

In order to understand the intuition behind this solution, one is
invited to consider a particular case where $h_L/h_H\approx 1$,
$\lambda=\mu=1$ and $\rho=1/2$. In this case, the region $B$ is
defined by the set of points where $|x_H-x_L|<\alpha/(1-\alpha)$.
Therefore, the set of points where the decision maker can use any
strategy depends strictly on the discount factor. If the discount
factor is large, the decision maker may keep queues with  a large
difference between their sizes. On the other hand, if the discount
factor is small this situation is not accepted as a solution
anymore.

\begin{figure}
\begin{picture}(100,100)(-30,0)
\thicklines \unitlength=0.6mm
\put(40,10){\vector(0,1){80}} \put(40,10){\vector(1,0){90}} 
 \put(135,10){$x_L$}
\put(30,95){$x_H$} \put(40,33){\line(3,1){75}}
\put(30,33){$\overline{\delta}$}\put(92,61){$x_H=(h_L/h_H)x_L+\overline{\delta}$}
\put(70,10){\line(3,1){75}}
\put(122,40){$x_H=(h_L/h_H)x_L+\underline{\delta}$}
\put(112,17){$A$} \put(80,25){$B$} \put(50,50){$C$}
\put(60,0){$-(h_H/h_L)\underline{\delta}$}
\end{picture}

\caption{The regions $A$, $B$ and $C$ in the plane $x_L - x_H$.}\label{fig0}
\end{figure}
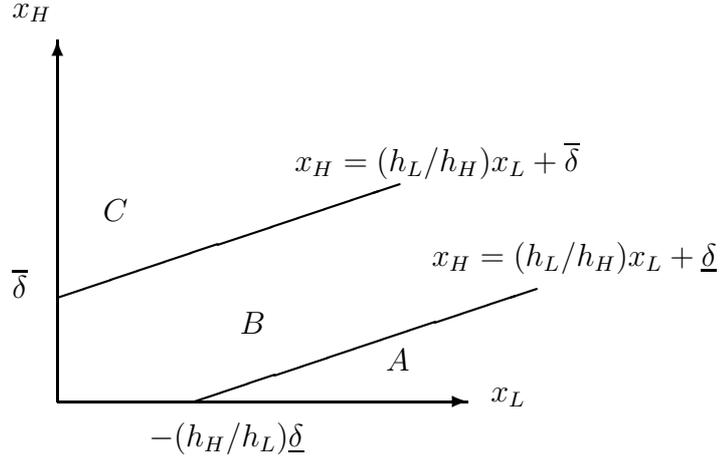

Differently from the linear costs case, several types of protocol
are possible. Region $C$ considers a protocol based on the
execution of the high priority task. Region $A$ considers a
protocol based on the execution of the low priority task. It
occurs in order to avoid that the size of the queue of the low
priority tasks do not increase too much. ``Too much'' here is
measured by the ratio $h_L/h_H$. Region $B$ does not determine a
protocol. It can be a random protocol (mixed strategy) or simply a
protocol such the one considered in region $C$ or region $A$.
Figure \ref{fig0} shows the geometry of these regions in the plane
$x_L - x_H$.

It is not difficult to show that the expected value of the state
obeys the following dynamics

\begin{eqnarray}E_t[x(t&+&1)]=E_t\left[\begin{array}{c} x_L(t+1) \\ x_H(t+1) \\\end{array}\right]=\left[\begin{array}{c} x_L(t) \\ x_H(t) \\\end{array}\right]\nonumber\\&+&\left[\begin{array}{c} \lambda (1-\rho)-\mu (1-u(x_L(t),x_H(t))) \\ \lambda \rho-\mu u(x_L(t),x_H(t)) \\\end{array}\right] \end{eqnarray}

which has infinite fixed points if and only if $\lambda=\mu$ and
$u(x_L,x_H)=u=\rho$.

We will analyze only the most interesting situation which is the
fixed-length-queue, i.e., $\lambda=\mu$. Therefore, assuming that
$\lambda=\mu$, $u^B=\rho+\epsilon$ and $\epsilon>0$, then the
expected value of the system is governed by
$E_t[x(t+1)]=x(t)+\lambda\epsilon e$ if it is in region $B$ and by
$E_t[x(t+1)]=x(t)-\lambda\rho e$ if it is in region $A$ (if the
state is in region $C$, the expected state will certainly come to
region $B$ and not come back to this region), where $e=(1,-1)'$.
Therefore, the dynamics takes place in the line passing by $x(0)$
and following the direction $e$. Thus, if the expected state is in
region $B$ it goes into the direction of region $A$ and viceversa.
This dynamics is equivalent to the one dimensional system

\begin{equation}y(t+1)=\left\{\begin{array}{c}
                                y(t)+t^{+}\;\mathrm{if}\; y(t)\le 0 \\
                                y(t)-t^{-}\;\mathrm{if}\; y(t)>0 \\
                              \end{array} \right. \label{dynamics}\end{equation}
defined on the interval $(-t^{-},t^{+}]$, where
$t^{+}=\lambda\epsilon$ and
$t^{-}=\lambda\rho$

\begin{figure}
\begin{center}
\includegraphics[width=8cm,height=8cm]{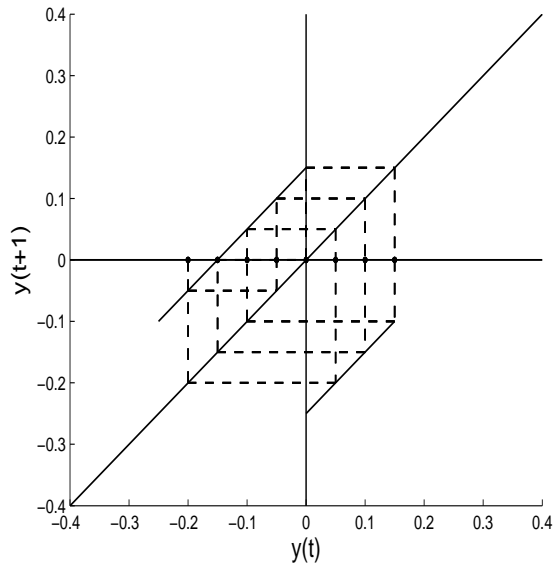}
\caption{The evolution of $y(t)$ for $y_0=-0.2$, $\lambda=0.5$,
$\rho=0.5$ and $\epsilon=0.3$.} \label{fig1}
\end{center}
\end{figure}

The dynamics of this system is plotted in figure \ref{fig1} for
the case of $t^{-}=0.25$ and $t^{+}=0.15$. The dynamics defined in
(\ref{dynamics}) is topologically conjugate with the translation
in the circle~\cite{dem93}. Therefore, if $t^+/t^-=p/q$, where
$p/q$ is a irreducible ratio representation of rational number,
then this system follows a limit cycle with period $p+q$.
Otherwise, the $\omega$-limit of any point in the interval is a
dense subset of it. Therefore, we can conclude that the stochastic
process that defines the length of each queue is not stationary.
Moreover, the dynamics of the expected value of the length of the
queue exhibits a complex behavior: infinitely many cycles or a
$\omega$-limit set being a dense subset in the interval. The
intuition behind that complex dynamics is quite reasonable. In the
region close to the frontier $x_H=(h_L/h_H)x_L
+\underline{\delta}$ that separate $A$ and $B$ (see figure
\ref{fig0}), we can observe the following: if the expected state
is in $A$, its dynamics moves toward region B, since the priority
is of $L$ . Once the expected state is in $B$, the dynamics takes
it back to the region A, since in this case in average the
priority is of $H$ (due to the condition $u^B=\rho+\epsilon$ and
$\epsilon>0$). Because the frequency of tasks arriving is equal to
that of attending them, a cyclical or complex dynamics emerges
close to the referred frontier. Figure \ref{fig1} shows the case
where this system is a limit cycle. A similar situation involving
regions $B$ and $C$ arises in the case of $\epsilon>0$ and
$\rho=u^B+\epsilon$. In these situations, the protocol is ruled by
the protocols considered in regions $A$ and $B$ in the former case
and by the protocols considered in regions $B$ and $C$ in the
later case.

For $\lambda\neq \mu$, either the expected value goes to infinite,
converges to 0, to axis $x_L=0$ or to axis $x_H=0$, following
different routes. Furthermore, different kinds of protocols
are possible.
\section{Final Remarks}In the human dynamics of the tasks execution decisions the priority of
one task is not always defined as being the most important current
task. Actually, the dynamics of the work executions depends on the
cumulated tasks of short run priorities, the importance of each
kind of task and the intertemporal discount factor. In this
letter, we provide a stochastic dynamic programming model
containing all those elements and analyze the dynamics of the
execution of tasks, shedding new light to the discursion
considered in~\cite{ken06,baroli06}. In this setting, we have
found that the dynamics of the expected state of the system may be
complex, exhibiting cycles of any order or with limit set being a
dense subset of the interval depending on the parameter values of
the model. This is a contribution to a better understanding of how
human dynamics may evolve in this type of problem. Finally, it is
worth noting that complex dynamics in the solution of dynamic
programming problems are usually obtained for low discount
factors~\cite{monsor96}. However, in our quadratic case, complex
dynamics arises for discount factors of any size.

\section{Acknowledgment} The authors are indebt to the Brazilian
agency CNPQ for financial support. The first author is also indebt
to Professor G. Grinstein (at IBM Watson Research Center) who
helped him by means of a personal communication to understand some
details of his paper~\cite{grilin06}.

\end{document}